# Towards 1000-mode Optical Fibres

Filipe M. Ferreira[(1)], Fabio A. Barbosa[(1)]

[(1)] Optical Networks, Dept. Electronic & Electrical Eng., University College London, f.ferreira@ucl.ac.uk

**Abstract** *We report on the design of multimode-mode fibres guiding up to 870 spatial and polarization modes for low differential mode delay over the C-band. ©2022 The Authors*

**Introduction**

Space-division multiplexing (SDM) has emerged as a solution to overcome the capacity limit of single-mode fibres (SMFs) [1]. Among the possible SDM approaches, multi-mode fibres (MMFs) offer the highest spatial information density followed by coupled-core multi-core fibres (MCFs) – and with bundles of SMFs and uncoupled-core MCFs on the other end.

Spatial density will play a critical role in maximising opportunities for opto-electronic integration gains, namely in transceivers, optical amplifiers, wavelength selective switches (WSSs), connectorisation and cable management and cable payload, and fibre-to-chip interface. SDM-specific transceiver integration is key to the overall value proposition of SDM. Spatial super-channels can share one laser for $N$ spatial tributaries (as opposed to combining $N$ lasers through a $1/N$ coupler – for WDM super-channels) and thus share common DSP functions such as laser frequency/phase recovery [2], beyond what can be achieved with parallel WDM systems. Also, with all crosstalking spatial paths originating and terminating at the same transceiver (as in coupled SDM), crosstalk cancellation techniques such as MIMO-DSP [3, 4] can be used, just as in wireless communications.

Multi-mode SDM offers the unique possibility of amplifying all spatial tributaries in a single Erbium-doped fibre pumped by a single high-power 980nm multimode pump with a wall-plug efficiency as much as 10 times better than that of single-mode version [5]. Also in switching, multimode SDM offers the unique possibility of having all spatial tributaries traversing the same optical elements as in a conventional WSS. This way offering great advantages over uncoupled SDM approaches that require WSS optics to handle many more free-space beams than switching ports [6].

SDM based on multi-mode fibres also presents major challenges. The multitude of spatial modes introduces new linear impairments, namely group delay (GD) spread [7-12] stemming from the interplay between differential mode delay (DMD) and linear mode coupling (LMC), and mode dependent loss (MDL) [4]. The GD spreading can be undone using of MIMO equalisation [3, 4] but with complexity scaling with the total time spread. Therefore, multimode SDM fibres are designed with a graded-index cores [13, 14] to reduce the DMD. However, higher the number of modes supported higher will be the minimum DMD achievable as it will be shown further on.

In this work, we aim to take first steps into understanding what would be the characteristics of optical fibres approaching 1000 spatial and polarisation modes with a DMD comparable to that of conventional multimode fibres. And what would be the trade-offs in attempting to do so.

**Methods**

It is well known that graded-index (GI) fibre cores are effective in reducing DMD as well as the two-fold benefits of using a cladding trench – enhanced confinement for reduced macro-bend loss and DMD reduction [13, 14].

Here we optimise GI profiles like the one in Fig. 1 for up to 870 spatial and polarization modes. Fig. 1 shows that there are up to 6 parameters passible of optimisation: $\alpha$, the core graded exponent, $\Delta n_{co}$ and $\Delta n_{tr}$, the core and trench relative refractive index, respectively, $w_2$, the trench to the core distance, and $w_3$, the trench width. However, a 6-D search space is not compatible with brute force searching. Therefore, the search space is reduced significantly using knowledge of the parameters interdependence and their relationship with the number of modes.

For each number of modes targeted ($M$), we choose the highest possible normalised frequency ($V = 2\pi w_1 / \lambda \cdot [n_{co}^2 - n_{cl}^2]^{1/2}$) [15] that guarantees the guidance of the first $M$ while cutting off the next higher-order modes. We followed the approach in [13] finding $V$ for $\alpha = 2$ and $\Delta n_{tr} = 0$. With $w_1$ bounded by with $\Delta n_{co}$ through normalised frequency choice, in the following we will only refer to $\Delta n_{co}$.

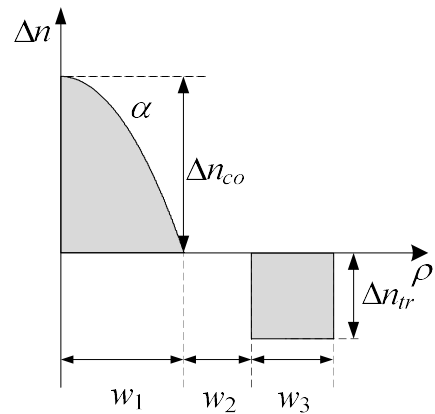

**Fig. 1:** GI profile with cladding trench

Regarding $\alpha$ and $\Delta n_{tr}$, the authors have previously shown that these two parameters define a convex space [13] greatly simplifying optimisation. Therefore, here, the search for the pair ($\alpha$, $\Delta n_{tr}$) that minimizes DMD for a given ($\Delta n_{co}$, $w_2$, $w_3$) is done one dimension at a time using for a golden section search (GSS).

The optimization function chosen is given by the maximum DMD (*maxDMD*) among the guided modes and over the wavelength range given of interest, in this case, the C-band:

$$maxDMD(pv) = max_\lambda(max_{\mu\nu}|DMD(\lambda, \mu\nu)|) \quad (1)$$

where $pv = [\alpha, \Delta n_{co}, \Delta n_{tr}, w_2, w_3]$, and DMD is the group delay difference between the slowest mode and the fastest mode.

Finally, the mode solutions of the waveguide defined by $pv$ are found using the vector finite difference mode solver developed in [16]. The dispersion properties of pure, Ge-doped or F-doped silica have been modelled using the Sellmeier coefficients provided in [17].

**Results**

During optimization the total diameter $2\cdot(w_1+w_2+w_3)$ is limited to be compatible with 125μm cladding fibres – for mechanical reliability equivalent to that of standard optical fibres. Interestingly, limiting the cladding diameter resulting in some fluorine cladding fibres being obtained as opposed to silica cladding fibres with a fluorine cladding trench.

In the following, the optimum profiles will be analysed not only for their intra and inter mode group DMD properties but also for linear mode coupling, chromatic dispersion, Rayleigh scattering and MBL. We will be working with linearly polarised (LP) waveguide solutions.

Fig. 2 shows *maxDMD* (as defined in (1)) as a function of the number of guided modes (targeted) for several $\Delta n_{co}$ values. It can be seen that higher $\Delta n_{co}$ allows supporting an increasingly larger number of modes (bounded by a 125μm cladding) but at the cost of a higher *maxDMD*. From $\Delta n_{co}\cdot 10^2$ with 11 mode groups, 66 LP modes, with *maxDMD* = 12 ps/km, all the way up to $\Delta n_{co}\cdot 10^2 = 3$ with 29 mode groups, 435 LP modes (i.e. 870 spatial and polarisation modes) with *maxDMD* = 554 ps/km. Tab. 1 summarises the results for other number of modes. Up to 276

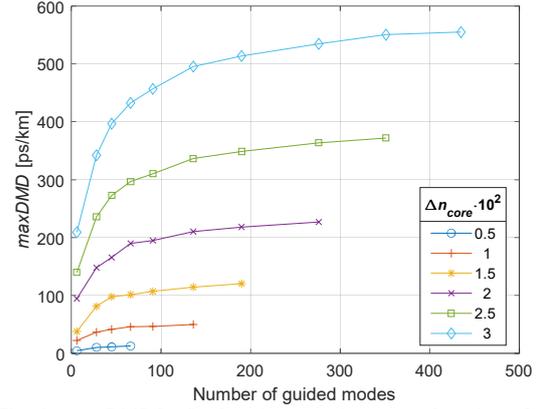

Fig. 2: *maxDMD* [ps/km] optimum values as a function of $\Delta n_{co}$ for different numbers of modes.

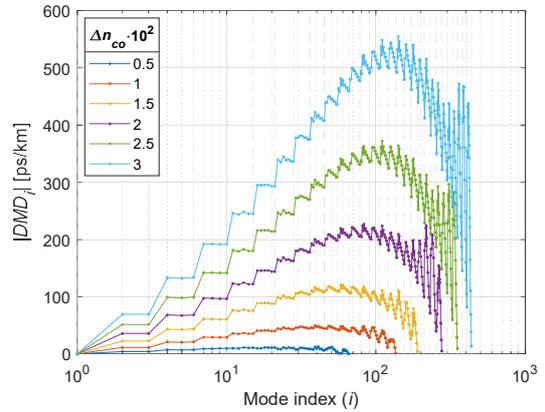

Fig. 3: $|DMD_i|$ [ps/km] as a function of the mode index $i$ for several $\Delta n_{co}$ – corresponding to the optimum fibre from Fig.5 carrying the highest number of modes for each $\Delta n_{co}$.

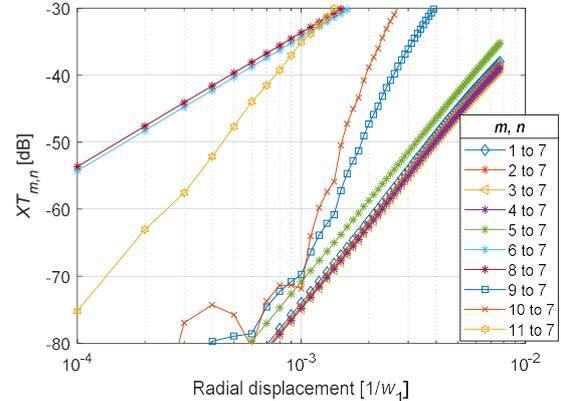

Fig. 4: $XT_{m,n}$ between mode group $m$ and $n$ as a function of the radial displacement for the optimum fibre guiding more modes with $\Delta n_{co} = 0.5$. $m$ taking over all other mode groups while $n = 7$.

LP modes DMD remains comparable to that of conventional FMFs and MMFs [18]. The practical impact of DMD on SDM capacity is well understood in terms of conventional MIMO equalisation complexity [3] – making the fibres in Fig. 2 candidates to data centre interconnect distances only. However, fibre link architectures inspired on those of (chromatic) dispersion-managed optical networks assisted by programmable optical equalisation [19-21] offer the possibility of extending the transmission reach.

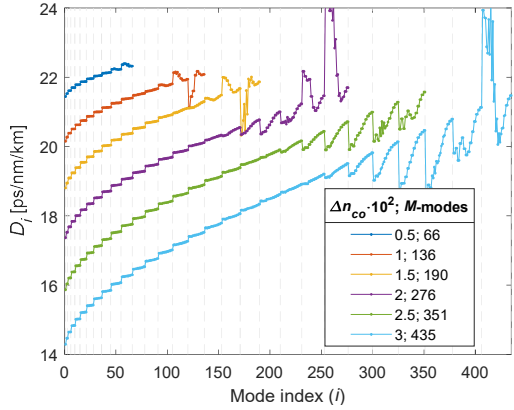

Fig. 5: *maxDMD* [ps/km] optimum values as a function of $\Delta n_{co}$ for different numbers of modes.

| *M*-modes | 66 | 136 | 190 | 276 | 351 | 435 |
|---|---|---|---|---|---|---|
| *maxDMD* | 12 | 49 | 120 | 226 | 372 | 554 |
| $\alpha_{Rayleigh}$* | .16 | .17 | .18 | .20 | .21 | .23 |
| | .17 | .20 | .23 | .26 | .29 | .32 |
| $\Delta n_{eff}$ 10³** | .6 | .9 | 1.1 | 1.3 | 1.4 | 1.6 |

Tab. 1: Optimum fibre characteristics as a function of *M*.

*for the fundamental and the highest order mode @ 1.55μm.
**diff. among neighbours (varies <2·10⁻⁴ for all groups) @ 1.55μm.

Fig. 3 shows the |*DMDi*| as a function of the mode index (*i*) for the optimum fibre with the largest number of modes for each $\Delta n_{co}$. It can be seen that within each mode group (highlighted by grey dashed vertical lines), the *DMD* varies within a few 1-10 ps/km up to a certain number of mode groups – the latter scaling with $\Delta n_{co}$. Low modal dispersion within mode groups offers the potential for mode group division multiplexing [20] – provided that leakage between neighbouring mode groups is not too strong (which depends on $\Delta n_{eff}$ [18]). Note that the fibre in Fig. 3 corresponding to $\Delta n_{co} \cdot 10^2 = 3$ guides over 200 modes with |*DMDi* − *DMDresp.MG*| lower than 100ps/km. Tab. 1 shows $\Delta n_{eff}$ between adjacent LP more groups. A large |$\Delta n_{eff}$| >1·10⁻³ (and even >1·10⁻³) is observed indicating that coupling could be stronger [18]. However, strong mode overlapping between adjacent mode groups might still significantly contribute to the overall crosstalk.

To gain insight into the mode coupling strength for the optimum fibres obtained, coupling is explicitly calculated for a single fibre imperfection consisting of lateral offset between two fibre segments. An approximate solution to the differential equations describing mode-coupling is used [22-24], $e^{j(\Delta\beta_{mn}+C_{m,n})\Delta z}$, where the mode-coupling coefficient is calculated as $C_{m,n} = \frac{\omega \varepsilon_0}{4} \int\int_{-\infty}^{+\infty} [\Delta\varepsilon(x,y)] E_m^* \cdot E_n dxdy$, with $\Delta\varepsilon(x,y)$ standing for the permittivity perturbation. Mode coupling *XTmn* among pair of mode groups *m* and *n* is calculated (from the approx. solution just described) considering only entries associated with *m* and *n*, and calculating the ratio between the power that remains on the launched modes and the unwanted modes. Mode coupling is evaluated for a fixed radial displacement by averaging over the azimuth displacement (with an arbitrarily small step). Fig. 4 shows *XT* as a function of the radial displacement. Besides within mode groups, LMC takes place mostly among neighbouring mode groups - unless for radial displacements larger than $w_1 \cdot 0.1\%$.

The macro bend loss (MBL) of the optimum fibres in Fig. 2 has also been evaluated. It has been found that MBL can be severe for some of modes in the last mode group of most fibres. Therefore, usage of the last two mode groups should be disregarded at this point. However, in future we plan to include MBL in the objective function to evaluate the trade-offs of overcoming the MBL in the last mode group – following [13].

Fig. 5 shows the chromatic dispersion per mode *i* for the optimum fibres in Fig. 2 carrying the highest number of modes for each $\Delta n_{co}$. The results compared well with that of typical standard SMF values (~17 ps/(nm·km)). Also chromatic dispersion is mostly constant within each mode group, expect for the last two mode groups of each fibre.

Further results have shown that Rayleigh scattering contribution to optical absorption scales as expected [25] from 0.16-0.17 dB/km for $\Delta n_{co} \cdot 10^2 = 0.5$ up to 0.23-0.32 dB/km for $\Delta n_{co} \cdot 10^2 = 3$, see Tab. 1 for other cases. Thus, $\Delta n_{co}$ allows to guide more modes not only at the expense of larger *DMD* but also of larger Rayleigh scattering due to the higher Ge concentration in the core.

## Conclusions

We have shown that multimode fibres can be designed to approach 1000 spatial and polarisation modes while keeping DMD comparable to that of conventional fibres. But we identified the scaling trend of several trade-offs such as those imposed by mode coupling, Rayleigh scattering and MBL. The next step will be to quantify how these limit information capacity and find the fundamental information spatial density limit for multimode fibres.

## Acknowledgements

This work was supported by the UKRI Future Leaders Fellowship MR/T041218/1. To access the underlying data for this publication, see: https://doi.org/10.5522/04/19739038.